\documentclass[prd,twocolumn,nofootinbib,superscriptaddress,amssymb,amsfonts,amsmath,preprintnumbers]{revtex4-2}
\usepackage{amsmath}
\usepackage{dsfont}
\usepackage{mathrsfs}
\usepackage{scalerel}
\usepackage[colorlinks=true,linktoc=page,linkcolor=purple,citecolor=teal,urlcolor=magenta]{hyperref}
\usepackage{url}
\usepackage{caption,subcaption,float}
\usepackage{graphicx,tabularx,tabularray,diagbox}
\usepackage[compat=1.1.0]{tikz-feynman}
\usepackage{xcolor}
\definecolor{dgreen}{rgb}{.0,.6,.0}
\definecolor{lime}{HTML}{A6CE39}
\definecolor{lg}{RGB}{220,220,220} 
\usepackage{ragged2e}
\usepackage{natbib}
\usepackage{cancel}

\makeatletter
\g@addto@macro\bfseries{\boldmath}
\makeatother	

\DeclareRobustCommand{\orcidicon}{\hspace{-2.1mm}
\begin{tikzpicture}
\draw[lime,fill=lime] (0,0.0) circle [radius=0.13] node[white] {{\fontfamily{qag}\selectfont \tiny ID}}; \draw[white,fill=white] (-0.0525,0.095) circle [radius=0.007]; 
\end{tikzpicture} \hspace{-3.7mm} }
\foreach \x in {A, ..., Z} {\expandafter\xdef\csname orcid\x\endcsname{\noexpand\href{https://orcid.org/\csname orcidauthor\x\endcsname} {\noexpand\orcidicon}}}

\newcommand{\Eprint}[1]{\href{#1}}

\newcommand\wh[1]{\hstretch{2}{\hat{\hstretch{.5}{#1\mkern1mu}}}\mkern-1mu}
\newcommand{\whbar}[1]{\widehat{\mskip.5\thinmuskip\overline{\mskip-.5\thinmuskip {#1} 
\mskip-.5\thinmuskip}\mskip.5\thinmuskip}}

\newenvironment{Eqnarray}%
     {\arraycolsep 0.14em\begin{eqnarray}}{\end{eqnarray}}
\def\beq{\begin{Eqnarray}}
\def\eeq{\end{Eqnarray}}
\def\beqa{\begin{Eqnarray}}
\def\eeqa{\end{Eqnarray}}
\def\bea{\begin{Eqnarray}}
\def\eea{\end{Eqnarray}}
\def\eq#1{Eq.~(\ref{#1})}

\def\Eqs#1#2{Eqs.~(\ref{#1}) and (\ref{#2})}
\def\eqs#1#2{Eqs.~(\ref{#1}) and (\ref{#2})}

\def\eqss#1#2#3{Eqs.~(\ref{#1}), (\ref{#2}), and (\ref{#3})}

\def\phm{\phantom{-}}
\def\pht{\phantom{i}}
\def\phaa{\phantom{AA}}

\def\ifmath#1{\relax\ifmmode #1\else $#1$\fi}
\def\lsub#1{\ifmath{_{\lower2.5pt\hbox{$\scriptstyle #1$}}}}
\def\lsup#1{^{\lower 2pt\hbox{$\scriptstyle#1$}}}
\let\Re\relax
\let\Im\relax
\DeclareMathOperator{\Re}{Re}
\DeclareMathOperator{\Im}{Im}

\def\half{\tfrac12}
\def\T{{\mathsf T}}
\def\nn{\nonumber}
\makeatletter
\let\save@mathaccent\mathaccent
\newcommand*\if@single[3]{%
  \setbox0\hbox{${\mathaccent"0362{#1}}^H$}%
  \setbox2\hbox{${\mathaccent"0362{\kern0pt#1}}^H$}%
  \ifdim\ht0=\ht2 #3\else #2\fi
  }
\newcommand*\rel@kern[1]{\kern#1\dimexpr\macc@kerna}
\newcommand*\widebar[1]{\@ifnextchar^{{\wide@bar{#1}{0}}}{\wide@bar{#1}{1}}}
\newcommand*\wide@bar[2]{\if@single{#1}{\wide@bar@{#1}{#2}{1}}{\wide@bar@{#1}{#2}{2}}}
\newcommand*\wide@bar@[3]{%
  \begingroup
  \def\mathaccent##1##2{%
    \let\mathaccent\save@mathaccent
    \if#32 \let\macc@nucleus\first@char \fi
    \setbox\z@\hbox{$\macc@style{\macc@nucleus}_{}$}%
    \setbox\tw@\hbox{$\macc@style{\macc@nucleus}{}_{}$}%
    \dimen@\wd\tw@
    \advance\dimen@-\wd\z@
    \divide\dimen@ 3
    \@tempdima\wd\tw@
    \advance\@tempdima-\scriptspace
    \divide\@tempdima 10
    \advance\dimen@-\@tempdima
    \ifdim\dimen@>\z@ \dimen@0pt\fi
    \rel@kern{0.6}\kern-\dimen@
    \if#31
      \overline{\rel@kern{-0.6}\kern\dimen@\macc@nucleus\rel@kern{0.4}\kern\dimen@}%
      \advance\dimen@0.4\dimexpr\macc@kerna
      \let\final@kern#2%
      \ifdim\dimen@<\z@ \let\final@kern1\fi
      \if\final@kern1 \kern-\dimen@\fi
    \else
      \overline{\rel@kern{-0.6}\kern\dimen@#1}%
    \fi
  }%
  \macc@depth\@ne
  \let\math@bgroup\@empty \let\math@egroup\macc@set@skewchar
  \mathsurround\z@ \frozen@everymath{\mathgroup\macc@group\relax}%
  \macc@set@skewchar\relax
  \let\mathaccentV\macc@nested@a
  \if#31
    \macc@nested@a\relax111{#1}%
  \else
    \def\gobble@till@marker##1\endmarker{}%
    \futurelet\first@char\gobble@till@marker#1\endmarker
    \ifcat\noexpand\first@char A\else
      \def\first@char{}%
    \fi
    \macc@nested@a\relax111{\first@char}%
  \fi
  \endgroup
}
\makeatother
\def\zv{\widebar{Z}_5}
\def\zvi{\widebar{Z}_6}
\def\zvii{\widebar{Z}_7}

\begin{document}

\preprint{PSI-PR-24-24, ZU-TH 58/24}

\title{Correlating $A\to \gamma\gamma$ with electric dipole moments in the two Higgs doublet \\ model in light of the diphoton excesses at 95\,GeV and 152\,GeV}

\author{Sumit Banik\orcidC{}}
\email{sumit.banik@psi.ch}
\affiliation{Physik-Institut, Universit\"{a}t Z\"{u}rich, Winterthurerstrasse 190, CH–8057 Z\"{u}rich, Switzerland}
\affiliation{Laboratory for Particle Physics, PSI Center for Neutron and Muon Sciences,
Forschungsstrasse 111, 5232 Villigen PSI, Switzerland}

\author{Guglielmo Coloretti\orcidA{}}
\email{guglielmo.coloretti@physik.uzH.c.h}
\affiliation{Physik-Institut, Universit\"{a}t Z\"{u}rich, Winterthurerstrasse 190, CH–8057 Z\"{u}rich, Switzerland}
\affiliation{Laboratory for Particle Physics, PSI Center for Neutron and Muon Sciences,
Forschungsstrasse 111, 5232 Villigen PSI, Switzerland}

\author{Andreas Crivellin\orcidB{}}
\email{andreas.crivellin@cern.ch}
\affiliation{Physik-Institut, Universit\"{a}t Z\"{u}rich, Winterthurerstrasse 190, CH–8057 Z\"{u}rich, Switzerland}
\affiliation{Laboratory for Particle Physics, PSI Center for Neutron and Muon Sciences,
Forschungsstrasse 111, 5232 Villigen PSI, Switzerland}

\author{Howard E. Haber\orcidD{\,\,}}
\email{haber@scipp.ucsc.edu}
\affiliation{Santa Cruz Institute for Particle Physics, University of California, 1156 High Street, Santa Cruz, CA 95064, USA}

\begin{abstract}
We examine the correlations between new scalar boson decays to photons and electric dipole moments (EDMs) in the CP-violating flavor-aligned two-Higgs-doublet model (2HDM). It is convenient to work in the Higgs basis $\{\mathcal{H}_1,\mathcal{H}_2\}$ where only the first Higgs doublet field $\mathcal{H}_1$ acquires a vacuum expectation value. In light of the LHC Higgs data, which agree well with Standard Model (SM) predictions, it follows that the parameters of the 2HDM are consistent with the Higgs alignment limit. In this parameter regime, the observed SM-like Higgs boson resides almost entirely in $\mathcal{H}_1$, and the other two physical neutral scalars, which reside almost entirely in $\mathcal{H}_2$, are approximate eigenstates of CP (denoted by the CP-even~$H$ and the CP-odd~$A$). In the Higgs basis, the scalar potential term $\zvii \mathcal{H}_1^\dagger \mathcal{H}_2 \mathcal{H}_2^\dagger \mathcal{H}_2+{\rm H.c.}$ governs the charged-Higgs loop contributions to the decay of $H$ and $A$ to photons.  If $\Re\zvii\Im\zvii\neq 0$, then CP-violating effects are present and allow for an $H^+ H^- A$ coupling, which can yield a sizable branching ratio for $A\to\gamma\gamma$. These CP-violating effects also generate nonzero EDMs for the electron, the neutron and the proton. We examine these correlations for the cases of $m_{A}=95$\,GeV and $m_{A}=152$\,GeV where interesting excesses in the diphoton spectrum have been observed at the LHC. These excesses can be explained via the decay of $A$ while being consistent with the experimental bound for the electron EDM in regions of parameter space that can be tested with future neutron and proton EDM measurements. This allows for the interesting possibility where the 95\,GeV diphoton excess can be identified with~$A$, while $m_H\simeq 98$\,GeV can account for the best fit to the LEP excess in $e^+e^-\to ZH$ with $H\to b\bar b$.
\end{abstract}
\maketitle

\section{Introduction}

With the discovery of the Higgs boson at the Large Hadron Collider (LHC)~\cite{Aad:2012tfa,Chatrchyan:2012ufa}, the Standard Model (SM) of particle physics is complete. Furthermore, the SM has been remarkably successful in describing the interactions of fundamental particles and their interactions (although there are a number of anomalies that, if verified in subsequent experimental studies, could provide definitive evidence for a breakdown of the SM~\cite{Crivellin:2023zui}).

Nevertheless, there is some motivation to extend the SM by adding additional scalar multiplets, as there is no consistency principle that requires the minimal realization of the Higgs sector in which one complex Higgs doublet with hypercharge $Y=\half$ generates mass for the $W^\pm$ and $Z$ gauge bosons, the quarks, and the charged leptons.
Indeed, in light of the observation that the SM possesses three generations of quarks and leptons, one might also expect additional generations of scalars (e.g., see Refs.~\cite{Lee:1973iz,Weinberg:1976hu,Georgi:1985nv,Gunion:1989we,Grossman:1994jb,Porto:2007ed,Ivanov:2017dad}).   Another indication of the need for additional scalars arises
when attempting to devise a theory of electroweak baryogenesis to explain the observed asymmetry between baryons and antibaryons (e.g., see Ref.~\cite{White:2022ufa}), which cannot be achieved by the SM alone. 

One of the most well-studied extensions of the SM scalar sector, which is obtained by adding a second complex Higgs doublet with hypercharge $Y=\half$, is the two-Higgs-doublet model (2HDM)~\cite{Lee:1973iz} (see Ref.~\cite{Branco:2011iw} for a comprehensive review).   The 2HDM provides opportunities for addressing the issues mentioned above while remaining compatible with electroweak precision data.

In its most general form, the 2HDM possesses new sources of CP violation due to additional complex parameters in the Higgs Lagrangian.  Such models can be compatible with a strong first-order electroweak phase transition needed for generating a sufficient baryon asymmetry in the early Universe~\cite{Sakharov:1967dj,Cohen:1993nk,Trodden:1998qg,Riotto:1999yt,Morrissey:2012db,FileviezPerez:2022ypk,Wagner:2023vqw}. However, in specialized versions of the 2HDM with natural flavor conservation, usually implemented via a $\mathbb{Z}_2$ symmetry~\cite{Hall:1981bc,Barger:1989fj,Aoki:2009ha}, there is only one additional physical complex phase in the scalar potential, such that achieving a large enough baryon asymmetry is challenging~\cite{Basler:2021kgq}. This difficulty can be overcome by giving up the $\mathbb{Z}_2$ symmetry and considering a more general 2HDM with additional sources of CP violation. 

However, new complex phases also give rise to contributions to low-energy probes of CP violation, in particular, electric dipole moments (EDMs)~\cite{Pospelov:2013sca,Yamaguchi:2020eub,Yamaguchi:2020dsy}. In the general CP-violating 2HDM, the EDM of the electron typically places the most stringent bound on the model parameters~\cite{Roussy:2022cmp,Abel:2020pzs,Higuchi:2022ihg}.  In some cases, future neutron and proton EDM measurements also have the potential to constrain the CP-violating parameters of the model.   

In the 2HDM with a SM-like Higgs boson (as required in light of LHC Higgs data~\cite{ATLAS:2022vkf,CMS:2022dwd}), there are two additional neutral scalars that are often approximate eigenstates of CP, denoted by the CP-even $H$ and the CP-odd $A$. Sizable branching ratios of $H$ and $A$ decays into photons are phenomenologically motivated by the $\gamma\gamma$ excesses at 95\,GeV~\cite{CMS:2018cyk,CMS:2022goy,CMS:2023boe,ATLAS:2024bjr} and 152\,GeV~\cite{ATLAS:2021jbf,ATLAS:2023omk,ATLAS:2024lhu,Crivellin:2021ubm,Bhattacharya:2023lmu,Coloretti:2023yyq,Ashanujjaman:2024pky,Crivellin:2024uhc,Fuks:2024qdt}.\footnote{The existence of a boson of mass $\sim$150\,GeV was first proposed in the context of the multilepton anomalies (see Refs.~\cite{Fischer:2021sqw,Crivellin:2023zui} for reviews) in $WW$ final states~\cite{vonBuddenbrock:2016rmr}, later found to be compatible with transverse mass~\cite{Coloretti:2023wng} differential top-quark distributions~\cite{Banik:2023vxa}.} In particular, it has been shown that it is difficult to achieve the rates preferred by data, especially if the CP-odd state is responsible for the excess at 95\,GeV or 152\,GeV within different versions of the 2HDM (see Refs.~\cite{Haisch:2017gql,Azevedo:2023zkg,Benbrik:2022azi,Benbrik:2024ptw} and Ref.~\cite{Banik:2024ftv}, respectively). In this paper, our aim is to broaden the treatment of the 2HDM version employed in analyzing these excesses, paying close attention to the correlations of the decay rate of $A\to \gamma\gamma$ with the EDMs of the electron, neutron, and proton.

In Sec.~\ref{formalism}, we review the most general CP-violating 2HDM, using the basis independent formalism introduced in Refs.~\cite{Davidson:2005cw,Haber:2006ue,Haber:2010bw,Boto:2020wyf}. In Sec.~\ref{decay}, the decay widths of the neutral Higgs bosons to two photons are obtained for the most general CP-violating 2HDM, and in Sec.~\ref{edm}, the electron, neutron and proton EDMs are considered.  In light of the EDM constraints, the viability of the general CP-violating 2HDM to explain the excesses in the $\gamma\gamma$ channel at 95\,GeV or 152\,GeV is examined in Sec.~\ref{sec::pheno}, under the assumption that one of the two experimental excesses represents new physics beyond the SM.
Brief conclusions are presented in Sec.~\ref{conclude}, followed by two appendices that provide details on the 2HDM Yukawa sector and summarize the loop functions employed in 
calculating the diphoton decays of the neutral scalars.

\section{2HDM Formalism}
\label{formalism}
The 2HDM employs two complex SU(2)$_L$ doublets scalars $\Phi_1$ and $\Phi_2$ with hypercharge $Y=\half$. In the most general version of the 2HDM, the fields $\Phi_1$ and $\Phi_2$ are indistinguishable. Thus, it is always possible to define a new basis of scalar fields, $\Phi^\prime _i=\sum_{j=1}^2 U_{ij}\Phi_j$ for $i=1,2$, where $U$ is a $2\times 2$ unitary matrix. In particular, one can always transform from the scalar field basis $\{\Phi_1,\Phi_2\}$ to the Higgs basis~\cite{Georgi:1978ri,Lavoura:1994yu,Lavoura:1994fv,Botella:1994cs,Branco:1999fs,Davidson:2005cw}, denoted by $\{\mathcal{H}_1, \mathcal{H}_2\}$, such that $\langle \mathcal{H}_1^0\rangle=v/\sqrt{2}$ and $\langle \mathcal{H}_2^0\rangle=0$, with $v\equiv
(\sqrt{2}G_F)^{-1/2}\simeq 246\,{\rm GeV}$, where $G_F$ is the Fermi constant.
Requiring renormalizability
and SU(2)$_L\times$U(1)$_Y$ gauge invariance, the most general scalar
potential in the Higgs basis is given by
 \beqa
 \hspace{-0.03in}\mathcal{V}&=& Y_1 \mathcal{H}_1^\dagger \mathcal{H}_1+ Y_2 \mathcal{H}_2^\dagger \mathcal{H}_2 +[Y_3 e^{-i\eta}
\mathcal{H}_1^\dagger \mathcal{H}_2+{\rm H.c.}] \nn \\[5pt]
&& +\,\half Z_1(\mathcal{H}_1^\dagger \mathcal{H}_1)^2+\half Z_2(\mathcal{H}_2^\dagger \mathcal{H}_2)^2
+Z_3(\mathcal{H}_1^\dagger \mathcal{H}_1)(\mathcal{H}_2^\dagger \mathcal{H}_2) \nn \\[3pt]
&&
+Z_4(\mathcal{H}_1^\dagger \mathcal{H}_2)(\mathcal{H}_2^\dagger \mathcal{H}_1)
+\left\{\half Z_5  e^{-2i\eta}(\mathcal{H}_1^\dagger \mathcal{H}_2)^2\right. \nn \\
&&   \left. \,\,\,\,+\big[Z_6  e^{-i\eta}\mathcal{H}_1^\dagger
\mathcal{H}_1 +Z_7 e^{-i\eta} \mathcal{H}_2^\dagger \mathcal{H}_2\big] \mathcal{H}_1^\dagger \mathcal{H}_2+{\rm
H.c.}\right\},
\label{higgsbasispot}
\eeqa
where $Y_1$, $Y_2$, and $Z_1,\ldots,Z_4$ are real, whereas $Y_3$, $Z_5$, $Z_6$, and $Z_7$ are (potentially) complex parameters. The minimization of the Higgs basis scalar potential yields 
\beq \label{hbasisminconds}
Y_1=-\half Z_1 v^2\,,\qquad\quad Y_3=-\half Z_6 v^2\,.
\eeq

The presence of the complex phase $e^{-i\eta}$ in \eq{higgsbasispot} accounts for the nonuniqueness of the Higgs basis, since one is always free to rephase $\mathcal{H}_2$ because its vacuum expectation value vanishes. In particular, under a U(2) basis transformation $\Phi_i\to U_{ij}\Phi_j$, the Higgs basis fields $\mathcal{H}_1$ and $\mathcal{H}_2$ are \textit{invariant} whereas the phase factor $e^{-i\eta}$ transforms as $e^{-i\eta}\to (\det~U) e^{-i\eta}$, where $\det U\equiv e^{i\phi}$ (such that $\phi\in\mathbb{R}$) is a complex number of unit modulus. It follows that the quantities $Y_1$, $Y_2$, and $Z_1,\ldots,Z_4$ are invariant with respect to a change of basis of the scalar fields, whereas $[Y_3, Z_6, Z_7]\to e^{-i\phi}[Y_3, Z_6, Z_7]$ and $Z_5\to  e^{-2i\phi} Z_5$.  Therefore, 
\beq \label{zvdef}
\zv\equiv Z_5 e^{-2i\eta}\,,\quad \zvi\equiv  Z_6 e^{-i\eta}\,,\quad  \zvii\equiv  Z_7 e^{-i\eta}\,,
\eeq
are basis-invariant quantities.\footnote{Note that physical observables can depend only on basis-invariant combinations of the scalar potential parameters.}

In the Higgs basis the scalar doublets 
can be parametrized as
\beqa
\label{hbasisfields}
&& \mathcal{H}_1=\left(\!\begin{array}{c}
G^+ \\[4pt] {\frac{1}{\sqrt{2}}}\left(v+\varphi_1^0+iG\right)\end{array}
\!\right),\quad\!\!
\mathcal{H}_2=\left(\!\begin{array}{c}
\mathcal{H}_2^+ \\[4pt] {\frac{1}{\sqrt{2}}}\left(\varphi_2^0+ia^0\right)\end{array}\!
\right),\nn \\
&& \phantom{line}
\eeqa
where we have identified $G=\sqrt{2}\,\Im\, \mathcal{H}_1^0$  and $G^\pm=\mathcal{H}_1^\pm$ 
as the massless CP-odd neutral and charged Goldstone bosons, respectively, and $\mathcal{H}_2^\pm$ as the physical charged Higgs boson pair with \beq \label{plusmass}
m_{H^\pm}^2=Y_2+\half Z_3 v^2\,.
\eeq
In general, the CP-even scalar $\varphi_1^0$ mixes with the scalars $\varphi_2^0$ and $a^0$.

The resulting physical neutral scalar
squared-mass matrix in the $\varphi_1^0$--$\varphi_2^0$--$a^0$ basis is
\beqa
&&\mathcal{M}^2=v^2\left( \begin{array}{ccc}
Z_1&\,\, \Re\zvi& -\Im\zvi\\[3pt]
\Re\zvi & \half\bigl[Z_{4c}+\Re\zv\bigr] & 
- \half \Im\zv \\[3pt]
 -\Im\zvi&  - \half \Im\zv& \half\bigl[Z_{4c}-\Re\zv\bigr]\end{array}\right), \nn \\
 && \phantom{line} \label{mtilmatrix}
\eeqa
where $Z_{4c}\equiv Z_4+2m_{H^\pm}^2/v^2$, and the quantities $\zv$ and $\zvi$ are defined in \eq{zvdef}.
If $\Im\zv=\Im(\zvi)^2=0$ then there is no mixing of the would-be CP-even and CP-odd neutral scalar states in the neutral scalar squared-mass matrix.  
If these conditions do not hold, then CP-violating interactions of the neutral scalar mass eigenstates are present.\footnote{Another potential source of
CP violation in the scalar self-interactions is due to the complex parameter $\zvii$, which 
does not appear in the mass matrix in Eq.~(\ref{mtilmatrix}) and is thus uncorrelated with the mixing among the neutral scalars. \label{fnz7}}

Since the squared-mass matrix $\mathcal{M}^2$ is real and symmetric, it can be diagonalized by an orthogonal transformation with unit determinant,
\beq \label{rmrt}
R\mathcal{M}^2 R^{\T}=\mathcal{M}^2_D\equiv {\rm diag}~(m_1^2\,,\,m_2^2\,,\,m_3^2)\,,
\eeq
where $RR^{\T}=I$, $\det R=1$, and the $m_k^2$ are the eigenvalues of $\mathcal{M}^2$.
A~convenient form for $R$ is
\beqa \label{rrrmatrix}
R&=&\left( \begin{array}{ccc}
c_{13}c_{12}\ &\,\, -s_{12}c_{23}-c_{12}s_{13}s_{23} &\,\, -c_{12}s_{13}c_{23}
+s_{12}s_{23}\\[3pt]
c_{13}s_{12}&\,\,  c_{12}c_{23}-s_{12}s_{13}s_{23}
&\,\, -s_{12}s_{13}c_{23}-c_{12}s_{23}\\[2pt]
s_{13}&\,\,  c_{13}s_{23}&\,\,  c_{13}c_{23}\end{array}\right), \nn \\
&&\phantom{line}
\eeqa
where $c_{ij}\equiv \cos\theta_{ij}$ and $s_{ij}\equiv\sin\theta_{ij}$.  
Indeed, the angles $\theta_{12}$, $\theta_{13}$, and $\theta_{23}$  defined above are basis-invariant quantities since they are obtained by diagonalizing $\mathcal{M}^2$, whose matrix elements are independent of the choice of the scalar field basis.

The neutral physical scalar mass eigenstates, denoted by $h_1$, $h_2$, and $h_3$ (with corresponding masses $m_1$, $m_2$, and $m_3$), 
are given by
\beqa \label{nrotated}
&&\left( \begin{array}{c}
h_1\\ h_2\\h_3 \end{array}\right)=R \left(\begin{array}{c} \varphi_1^0\\
\varphi_2^0\\ a^0 \end{array}\right)=Q \left(\begin{array}{c}
\sqrt{2}\,\Re~\mathcal{H}_1^0-v\\ \mathcal{H}_2^0\\  \,\,\mathcal{H}_2^{0\,\dagger} \end{array}\right)\,,
\eeqa
where 
\beqa
Q= \left(\begin{array}{ccc}q_{11} 
&\,\,\,  \tfrac{1}{\sqrt{2}}q^*_{12}\,e^{i\theta_{23}}
&\,\,\,  \tfrac{1}{\sqrt{2}} q_{12}\,e^{-i\theta_{23}}\\[5pt]
q_{21} &\,\,\,   \tfrac{1}{\sqrt{2}}q^*_{22}\,e^{i\theta_{23}}
&\,\,\,  \tfrac{1}{\sqrt{2}}q_{22}\,e^{-i\theta_{23}} \\[5pt]
q_{31} &\,\,\,  \tfrac{1}{\sqrt{2}}q^*_{32}\,e^{i\theta_{23}}
&\,\,\,  \tfrac{1}{\sqrt{2}} q_{32}\,e^{-i\theta_{23}}
\end{array}\right),
\eeqa
and the $q_{kj}$ are defined in Table~\ref{tabqij}. 

\begin{table}[t!]
\centering
\begin{tabular}{|c||c|c|}\hline
$\phaa k\phaa $ &\phaa $q_{k1}\phaa $ & \phaa $q_{k2} \phaa $ \\ \hline
$1$ & $c_{12} c_{13}$ & $-s_{12}-ic_{12}s_{13}$ \\
$2$ & $s_{12} c_{13}$ & $c_{12}-is_{12}s_{13}$ \\
$3$ & $s_{13}$ & $ic_{13}$ \\
\hline
\end{tabular}
\caption{\small \justifying The basis-invariant quantities $q_{k\ell}$ are functions of the neutral scalar 
mixing angles $\theta_{12}$ and $\theta_{13}$, with
$c_{ij}\equiv\cos\theta_{ij}$ and $s_{ij}\equiv\sin\theta_{ij}$. The angles $\theta_{12}$, $\theta_{23}$ are defined modulo $\pi$.
By convention, we take $0\leq c_{12}, c_{13}\leq 1$.\phantom{xxxxxxxxxxxxxxx}
\label{tabqij}}
\end{table}

It is convenient to define the positively charged Higgs field as
\beq \label{chhiggs}
H^+ \equiv e^{i{\theta}_{23}}\mathcal{H}_2^+\,.
\eeq
One can then invert \eq{nrotated} and include the charged scalars to obtain
\beqa
\mathcal{H}_1&=&\begin{pmatrix} G^+ \\[3pt] \displaystyle\frac{1}{\sqrt{2}}\left(v+iG+\sum_{k=1}^3 q_{k1}h_k\right)\end{pmatrix},  \label{Hbasismassbasis1}
\\[6pt]
e^{i{\theta}_{23}}\mathcal{H}_2&=&\begin{pmatrix} H^+ \\[3pt] \displaystyle\frac{1}{\sqrt{2}}\sum_{k=1}^3 q_{k2}h_k\end{pmatrix}.   \label{Hbasismassbasis2}
\eeqa
Although ${\theta}_{23}$ is a basis invariant parameter, it has no physical significance since it can be eliminated by rephasing the scalar doublet field $\mathcal{H}_2\to e^{-i\theta_{23}}\mathcal{H}_2$.  Henceforth, we shall set $\theta_{23}=0$ as advocated in Refs.~\cite{Haber:2022gsn,Connell:2023jqq}.

To perform a phenomenological analysis, it is instructive to identify a set of independent basis-invariant parameters that govern the scalar sector of the most general 2HDM.  First, we 
rewrite \eq{rmrt} as $\mathcal{M}^2=R^{\T}\mathcal{M}^2_DR$ and insert the expression for $R$ given by \eq{rrrmatrix} with $\theta_{23}=0$.  We then make use of 
\eq{mtilmatrix} to obtain
\beqa
Z_1  &=& \frac{1}{v^2}\sum_{k=1}^3 m_k^2 (q_{k1})^2\,,\label{zee1id} \\
Z_4  &=& \frac{1}{v^2}\left[\,\sum_{k=1}^3 m_k^2 |q_{k2}|^2 -2m_{H^\pm}^2\right]\,, \label{zeefour}  \\
\zv &=& \frac{1}{v^2}\sum_{k=1}^3 m_k^2 (q_{k2}^*)^2\,, \label{zee5id}\\
\zvi &=& \frac{1}{v^2}\sum_{k=1}^3 m_k^2 \,q_{k1} q_{k2}^*\,.\label{zee6id}
\eeqa
Together with Eqs.~(\ref{hbasisminconds}) and (\ref{plusmass}) it follows that a convenient choice for the set of independent parameters is
\beq
\{v,\theta_{12},\theta_{13},m_1,m_2,m_3,m_{H^\pm},Z_2,Z_3,\Re\zvii,\Im\zvii\}\,,\phantom{xxx}
\eeq
where the complex parameter $\zvii$ is defined in \eq{zvdef}. That is, the scalar sector of the most general 2HDM is governed by 11 real independent basis-invariant parameters.

The LHC Higgs data strongly suggest that the observed Higgs scalar of mass 125\,GeV is SM-like~\cite{ATLAS:2022vkf,CMS:2022dwd}.  We shall identify this scalar with $h_1\simeq h_{\rm SM}$. As the $W^+W^- h_k$ terms in the Lagrangian are given by
\beq
W^+ _\mu W^-_\nu h_k:\qquad\qquad gm_W q_{k1}g_{\mu\nu}\,,\phantom{xxxx}
\eeq
for $k=1,2,3$, it follows that the $h_1 W^+ W^-$ coupling coincides with that of the SM Higgs boson if $q_{11}=1$, which corresponds to the Higgs alignment limit~\cite{Ginzburg:2001wj,Gunion:2002zf,Craig:2013hca,Asner:2013psa,Carena:2013ooa,Haber:2013mia}. In this limit, $c_{12}=c_{13}=1$ and $s_{12}=s_{13}=0$ (or equivalently,  $q_{11}=q_{22}=-iq_{32}=1$ and $q_{21}=q_{31}=q_{12}=0$ in light of Table~\ref{tabqij}). 
Using \eqs{zee5id}{zee6id}, one obtains
\beq \label{alignconds}
\Im\zv=\zvi=0\,,\quad \text{(Higgs alignment limit)}\,.
\eeq

The conditions for a CP-conserving scalar sector are given by~\cite{Davidson:2005cw}: 
\beq \label{cpvconds}
\Im(\zv^* \zvi^2)=\Im(\zv^* \zvii^2)=\Im(\zvi^* \zvii)=0\,.
\eeq
Whereas
$\zvii$ does not appear in the neutral scalar squared-mass matrix, it provides a potentially new source of CP-violation via the scalar self-interactions. For example, the $H^+H^-h_k$ terms that appear in the scalar Lagrangian, obtained after inserting \eqs{Hbasismassbasis1}{Hbasismassbasis2} into \eq{higgsbasispot}, are given by 
\beq \label{count}
H^+ H^- h_k:\qquad -v\bigl[q_{k1}Z_3+\Re(q_{k2}\zvii)\bigr]\,,
\eeq
with $k=1,2,3$. 
These interactions are relevant for the charged Higgs boson loop contribution to the $h_k\to\gamma\gamma$ decay amplitude because they give unsuppressed contributions in the Higgs alignment limit. Therefore we have used them in Sec.~\ref{decay} to generate a sizable branching ratio of $h_2$ and $h_3$ to two photons.

As a result of \eqs{alignconds}{cpvconds}, it follows that the scalar sector is CP-conserving if and only if, 
\beq
\Im(\zv^* \zvii^2)=2\zv\Re\zvii\,\Im\zvii=0\,.
\eeq
Under the assumption of $m_2\neq m_3$, $\zv$ is real and nonzero [see \eq{zee5id}], in which case CP conservation requires that either $\Re\zvii=0$ or $\Im\zvii=0$. 
In the Higgs alignment limit, the $H^+ H^- h_k$ couplings are given by
\beqa \label{align}
&& H^+ H^- h_1:\qquad -v Z_3\,, \\
&& H^+ H^- h_2:\qquad -v\Re \zvii\,, \label{HpHmH} \\
&& H^+ H^- h_3:\qquad  \phm v\Im \zvii\,.\label{HpHmA}
\eeqa
Since $H^+ H^-$ is CP-even, it follows that $h_1$ is CP-even in the Higgs alignment limit, whereas $h_2$ is CP-even and $h_3$ is CP-odd if $\Im\zvii=0$.\footnote{Note that if $\Re\zvii=0$ then 
$h_2$ is CP-odd and $h_3$ is CP-even in the Higgs alignment limit.  That is, the CP properties of $h_2$ and $h_3$ are reversed with respect to the case of $\Im\zvii=0$.  
If $\zvii=0$, then the CP quantum numbers of $h_2$ and $h_3$ are not individually determined even though the bosonic sector of the theory is CP-conserving and the presence of the $Zh_2 h_3$ coupling indicates that the product of fields $h_2 h_3$ is CP-odd.} In contrast, if $\Re\zvii\,\Im\zvii\neq 0$, then $h_2$ and $h_3$ are states of indefinite CP and the scalar sector is CP-violating.

In practice the Higgs alignment limit of the 2HDM is not exact. Indeed, the LHC Higgs data (which confirms that $h_1$ is SM-like) only requires that $|s_{12}|$, $|s_{13}|\ll 1$.  Consider the case of $0<|s_{12}|\ll 1$ and $s_{13}=0$.  Then \eqs{zee5id}{zee6id} yield $\Im \zv=\Im\zvi=0$. If in addition $\zvii=0$, then the bosonic sector is CP-conserving and we can identify $h_2\equiv H$ and $h_3\equiv A$, where we have employed the standard notation of the 2HDM where $H$ is CP-even and $A$ is CP-odd.  
It is then useful to maintain the $H$, $A$ notation even when $\zvii\neq 0$, in which case, an $H^+H^- A$ coupling (which violates CP) will be present if $\Im\zvii\neq 0$ as indicated in \eq{HpHmA}. Similarly, if $0<|s_{13}|\ll 1$ and $s_{12}=0$, then it follows that $\Im \zv=\Re\zvi=0$, in which case we can identify $h_2\equiv A$ and $h_3\equiv H$.  In this scenario, an $H^+H^- A$ coupling (which violates CP) will be present if $\Re\zvii\neq 0$ as indicated in \eq{HpHmH}.

Finally, we examine the Higgs-fermion couplings of the 2HDM. In the absence of a $\mathbb{Z}_2$ symmetry to constrain the Higgs Lagrangian, both scalar doublets will couple to up-type and down-type fermions in the Yukawa Lagrangian, 
\beqa
-\mathscr{L}_Y&=& \overline{Q}(\boldsymbol{\wh{\kappa}_{U}} \widetilde{\mathcal{H}}_1
 +  \boldsymbol{\wh{\rho}_{U}}  \widetilde{\mathcal{H}}_2\,) U  +  \overline{Q}(\boldsymbol{\wh{\kappa}_D^{\dagger}} {\mathcal{H}}_1
 +   \boldsymbol{\wh{\rho}_D^{\dagger}} {\mathcal{H}}_2\,) D\nn \\
&& + \overline{L}( \boldsymbol{\wh{\kappa}_E^{\dagger}}{\mathcal{H}}_1
 +   \boldsymbol{\wh{\rho}_E^{\dagger}} {\mathcal{H}}_2\,) E
+{\rm H.c.},\label{yukawas}
\eeqa
where $\overline{Q}\widetilde{\mathcal{H}}_k\equiv\overline{Q}\lsup{1} \mathcal{H}^{2\dagger}_k-\overline{Q}\lsup{2} \mathcal{H}^{1\dagger}_k$, $\overline{Q}\mathcal{H}_k\equiv\overline{Q}\lsup{i}\mathcal{H}^{i}_k$,
and $\overline{L}\mathcal{H}_k\equiv\overline{L}\lsup{i} \mathcal{H}^{i}_k$, summed over the repeated SU(2)$_L$ superscript index~$i$, for $k=1,2$. 
In \eq{yukawas},  $\boldsymbol{\wh{\kappa}_{F}}$,  $\boldsymbol{\wh{\rho}_{F}}$ (for $F=U,D,E$)
are $3\times 3$ Yukawa coupling matrices, $Q$ and $L$ are SU(2)$_L$ doublets of left-handed quark and lepton fields, and $U$, $D$ and $E$ are SU(2)$_L$ singlets of right-handed quark and lepton fields (with the generation index suppressed). When the fermion mass matrices, $\boldsymbol{\wh{M}_F}\equiv v\boldsymbol{\wh{\kappa}_F}/\sqrt{2}$, 
are diagonalized, the corresponding transformed $\boldsymbol{\wh{\rho}_F}$ matrices are in general complex and nondiagonal. That is, without further restrictions on the Yukawa Lagrangian, natural flavor conservation cannot be enforced~\cite{Glashow:1976nt,Paschos:1976ay}.  As a result, the most general 2HDM will generically yield dangerously large flavor-changing neutral currents at tree-level mediated by neutral scalars (e.g., see Ref.~\cite{Crivellin:2013wna} for a detailed analysis).

An alternative approach to avoid off-diagonal neutral scalar couplings to fermions is to impose alignment in flavor space on the Yukawa couplings of the two scalar doublets~\cite{Pich:2009sp,Pich:2010ic,Eberhardt:2020dat,Choi:2020uum,Lee:2021oaj,Serodio:2011hg,Cree:2011uy,Connell:2023jqq}.  That is, we define (potentially complex) flavor alignment parameters $a_F$ in \eq{yukawas} via:\footnote{Flavor-aligned extended Higgs sectors can arise naturally from symmetries of ultraviolet completions of low-energy effective theories of flavor as shown in Refs.~\cite{Serodio:2011hg,Knapen:2015hia,Egana-Ugrinovic:2018znw,Egana-Ugrinovic:2019dqu}.   In such models, departures from exact flavor alignment due to renormalization group running down to the electroweak scale are typically small enough~\cite{Braeuninger:2010td,Gori:2017qwg} to be consistent with all known experimental FCNC bounds.}
\beq \label{alignedparms}
\boldsymbol{\wh{\rho}_F}=\frac{\sqrt{2}}{v} a_F \boldsymbol{\wh{M}_F}\,,\qquad\quad \text{for $F=U,D,E$}.
\eeq
Note that the Yukawa coupling matrices defined in \eq{yukawas} and the flavor alignment parameters $a_F$ are invariant with respect to a change of basis of the scalar fields.

In Appendix~\ref{app::Yukawa}, we obtain the Yukawa Lagrangian involving the neutral scalar fields $h_k$ [cf.~\eq{YUK4}]:
\beqa
&& -\mathscr{L}_Y = \frac{1}{v}\,\overline U \boldsymbol{M_U}\sum_{k=1}^3 \bigl(q_{k1}
+ q_{k2}^* a_U \mathcal{P}_R+q_{k2} a_U^*\mathcal{P}_L\bigr)Uh_k   \nonumber \\
&& +\frac{1}{v}\sum_{F=D,E} \overline F  \boldsymbol{M_F}\sum_{k=1}^3 \bigl(q_{k1}
+ q_{k2} a_F^* \mathcal{P}_R+q^*_{k2}a_{F}\mathcal{P}_L\bigr) Fh_k\,. \nn \\
&&\phantom{line}
\label{YUKneutrals}
\eeqa
In the exact Higgs alignment limit, \eq{YUKneutrals} reduces to
\beqa
&& \hspace{-0.3in} -\mathscr{L}_Y = \frac{1}{v}\sum_{F=U,D,E} \overline{F}\boldsymbol{M_F}Fh_1 \nn \\
&&+\frac{1}{v}\sum_{F=U,D,E}\overline{F}\boldsymbol{M_F}\bigl(\Re a_F+i\varepsilon_F\gamma\lsub{5}\Im a_F\bigr)Fh_2 \nn \\
&&+\frac{1}{v}\sum_{F=U,D,E}\overline{F}\boldsymbol{{M}_F}\bigl(\Im a_F-i\varepsilon_F\gamma\lsub{5}\Re a_F\bigr)Fh_3\,,\label{YUK5}
\eeqa
where we have introduced the notation
\beq \label{epsilonF}
\varepsilon_F\equiv \begin{cases} +1\,, & \text{for $F=U$}, \\ -1\,, & \text{for $F=D,E$}. \end{cases}
\eeq
Note that the Yukawa Lagrangian exhibited in \eq{YUK5} is CP conserving if the alignment parameters $a_U$, $a_D$, and $a_E$
are either all real (thereby identifying $h_2=H$ and $h_3=A$) or all pure imaginary (where the corresponding CP properties of $h_2$ and $h_3$ are reversed).

\section{Neutral scalar decay to photons}
\label{decay}

The diphoton partial decay widths of the neutral scalars are induced at one-loop by diagrams involving $W$ bosons, charged scalars, quarks, and charged leptons. Using the formulas given in Refs.~\cite{Gunion:1989we},
\begin{widetext}
\beqa \label{eq::saa}
    \Gamma(h_k \to \gamma \gamma) &=& \!\frac{G_F \alpha^2 m_k^3}{128 \pi^3\sqrt{2} } \left\{\biggl|q_{k1} A_W(\tau_W) + \frac{v^2}{2 m^2_{H^\pm}}\bigl[q_{k1}Z_3+\Re(q_{k2}\zvii)\bigr]A_{H^\pm}(\tau_{H^\pm}) +  \sum_f N_{cf}e_f^2 \bigl[q_{k1}+\Re(q_{k2}^*a_f)\bigr] A_f^{0}(\tau_f) \biggr|^2\right. \nn \\
 && \qquad\qquad\quad  \left. + \biggl| \sum_f N_{cf} e_f^2 \Im(q_{k2}^*a_f) A_f^{5}(\tau_f) \biggr|^2 \right\},
\eeqa
\end{widetext}
where the sum over $f$ is taken over three generations of up-type quarks (with $a_f=a_U$), down-type quarks (with $a_f=a_D$), and charged leptons (with $a_f=a_E$), the
$e_f$ are the corresponding fermion charges in units of the positron charge $e$, $N_{cf}=3$  ($N_{cf}=1$)  for the quarks (leptons), the loop functions $A_W$, $A_{H^\pm}$, $A_f^0$ and $A_f^5$ are listed in Appendix~\ref{app::LoopA}, and
$\tau_X\equiv 4m_X^2/m_k^2$.   We cross-checked our results with $\texttt{ScannerS}$~\cite{Muhlleitner:2020wwk}, confirming our results for the case $\lambda_6=\lambda_7=0$ in a generic basis of scalar fields, and assuming the Yukawa sector of the type-I 2HDM~\cite{Fontes:2017zfn}.

Note that for exact Higgs alignment, the partial decay width $\Gamma(h_1\to\gamma\gamma)$ differs from its SM value by the contribution of the charged Higgs loop that is proportional to $(Z_3)^2$. The contributions of the fermion loops to the diphoton partial decay widths of $h_2\simeq H$ and $h_3\simeq A$ vanish in the limit of $a_F=0$. In this approximation, the diphoton partial decay widths of $H$ and $A$ arise solely from the charged Higgs loop, with contributions proportional to $[\Re\zvii]^2$ and $[\Im\zvii]^2$, respectively.

The partial decays widths of the neutral scalars to fermions, $WW$ and $ZZ$ can be obtained by rescaling the one of a hypothetical SM Higgs boson with the same mass~\cite{LHCHiggsCrossSectionWorkingGroup:2016ypw} by the square of the ratios of the corresponding vertex factors.\footnote{For CP-eigenstates, the Higgs decays widths to light fermion pairs are roughly the same for a CP-even and a CP-odd scalar, whereas the latter has no couplings to gauge bosons~\cite{Gunion:1989we}. For the case of mixed states, the CP-even and CP-odd scalar couplings to fermions do not interfere when calculating the decay widths and only the CP-even component of a scalar couples to $WW$ and $ZZ$. Note that complex couplings also lead to the decays $H^\pm\to W^\pm Z$; however with very small branching ratios~\cite{Kanemura:2024ezz}.} In this way higher-order QCD and electroweak corrections are included.

\section{Electric Dipole Moments}
\label{edm}

The most constraining bound on the CP-violating parameters of the general 2HDM derives from the measurement of the electron EDM~\cite{Roussy:2022cmp,Abel:2020pzs,Higuchi:2022ihg},
\begin{align}
    |d_e| & \leq (1.3 \pm 2.0_{\text{stat}} \pm 0.6_{\text{sys}})\times 10^{-30}~e\,\text{cm}.
\end{align}
For calculating the electron EDM within the 2HDM we used the results and the publicly available code provided as supplemental material in Ref.~\cite{Altmannshofer:2020shb} (see~Eq.~(41) and related discussion in Ref.~\cite{Altmannshofer:2020shb}).  As in Sec.~\ref{decay}, we again have used \texttt{ScannerS}~\cite{Muhlleitner:2020wwk} to cross-check our results for the electron EDM. 

\begin{figure}[ht] 
            \begin{subfigure}{0.45\textwidth} \label{fig::EDMgeneric}
            \includegraphics[]{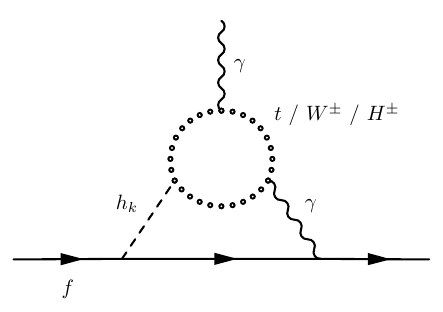}
            \caption{}
            \end{subfigure} 
            \begin{subfigure}{0.45\textwidth} \label{fig::EDMweinberg}
            \includegraphics[]{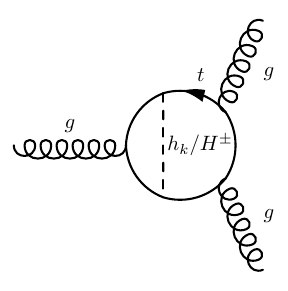}
            \caption{}
            \end{subfigure}
            \caption{ \justifying \small Representative Feynman diagrams contributing to EDMs: (a)  example of a Barr-Zee diagram that contributes to fermionic EDMs. Note that if a $t$-quark is in the loop, then a chromomagnetic operator is also induced by replacing both photons with gluons; (b) contribution to the Wilson coefficient of the three-gluon Weinberg operator. }
            \label{fig::EDMs}
\end{figure}

The prospects for the neutron EDM ($|d_n|$) and proton EDM ($|d_p|$) are at the level of $10^{-27}\,e\,\text{cm}$~\cite{Higuchi:2022ihg} and $10^{-29}\,e\,\text{cm}$~\cite{Omarov:2020kws}, respectively. The dominant contributions in the 2HDM arise at the two-loop level via the Barr-Zee diagrams~\cite{Barr:1990vd} as well as sunset diagrams contributing to the three-gluon Weinberg operator\footnote{There is also another contribution to the three-gluon Weinberg operator at 2-loops, the so-called charged contribution corresponding to the lower Feynman diagram in Fig.~\ref{fig::EDMs} with the charged Higgs boson in the propagator inside the loop. However, this contribution is proportional to $\Im(a_D^* a_U)$ and can be safely neglected in our numerical analysis.} $\tilde{d}_G(m_{t})$~\cite{Weinberg:1989dx} (see Fig.~\ref{fig::EDMs}). The neutron EDM can then be written as~\cite{Alarcon:2022ero,Bhattacharya:2021lol}
\begin{align} \label{nEDM}
     \nn d_n =& - (0.20 \pm 0.01)d_u + (0.78 \pm 0.03)d_d \\
     \nn & -(0.55 \pm 0.28) e \tilde{d}_u -(1.1 \pm 0.55)e\tilde{d}_d \\
     &+  (50\pm40)~\text{MeV}~e\tilde{d}_G\,,
\end{align}
where the theory errors associated with the hadronic matrix elements~\cite{Weinberg:1989dx,Yamanaka:2020kjo,Dragos:2019oxn,Gupta:2018lvp,Bhattacharya:2015esa,Pospelov:2000bw,Hisano:2012sc} have been neglected.  In \eq{nEDM}, the chromomagnetic contributions have been labeled with a tilde and~\cite{Jung:2013hka,Teunissen:2023lvt,Kuramoto:2017mdg}
\beqa
&&   \tilde{d}_G(m_{t})= 4 g_s^3(m_t)\frac{\sqrt{2}G_F}{(4\pi)^4} \nn \\
&& \quad \times \sum_{f=t,b} \varepsilon_f [q_{k1}+\Re(q_{k2}^*a_f)]\Im(q_{k2}^*a_f)F(m_f,m_k),\phantom{xxxx}
\eeqa
where $\varepsilon_f=\pm1$ for $f=t$ (with $a_f=a_U$) and $f=b$ (with $a_f=a_D$) [cf.~\eq{epsilonF}], and $F(m,M)$ is defined as
\begin{align}
    \nn &F(m_f,m_k)=\\
    &\frac{m_f^4}{4}\int^1_0 dx\int^1_0 du \frac{u^3x^3(1-x)}{[m_f^2 x(1-ux) + m_k^2(1-u)(1-x)]^2}\,.
\end{align}
The contributions of light quarks were determined using the prescription described in Eq.~(63) of Ref.~\cite{Altmannshofer:2020shb} for recasting the corresponding results of the electron. We used renormalization-group-improved results for the chromomagnetic contribution~\cite{Jung:2013hka,Dekens:2018bci}. To obtain the proton EDM, to first approximation one can swap down quarks with up quarks in Eq.~(\ref{nEDM}).

\section{Phenomenology} \label{sec::pheno}

As discussed in Sec.~\ref{edm}, $\Im\zvii$ is the only scalar potential parameter that generates an unsuppressed contribution to the decay width $(h_3\simeq A)\to\gamma\gamma$. If, in addition, the mixing with other scalars and the flavor-alignment parameters $a_F$ ($F=U$, $D$, $E$) are small, then it is possible to enhance the branching ratio, BR$(h_3\to\gamma\gamma)$, beyond a value that is accessible in a 2HDM with a $\mathbb{Z}_2$ conserving scalar potential. Let us consider \textit{separately} the diphoton excesses at 95\,GeV and 152\,GeV as applications for this mechanism.

\subsection{95\,GeV and 98\,GeV excesses} \label{ninefive}
\captionsetup[table]{
  position=above, aboveskip=4pt,
  belowskip=-4pt
}
\begin{table*}
\begin{center}
\SetTblrInner{colsep=.5pt}
\begin{tblr}{c|c|c|c|c|c|c|c|c|c|c|c}
	&	$b\bar b$		& 	$\tau^+\tau^-$	&	$c\bar c$	&	$\mu^+ \mu^-$	&	$W^+ W^-$	&	$ZZ$	 	&	$gg$		&	$\gamma\gamma$	& \pht $Zh_1$ \pht &\pht $Zh_i$ \pht	& \pht	$W^\pm H^\mp$\\
\hline
$h_2 \simeq H$\pht &\pht $0.80$ \pht &\pht  $0.084$ \pht & \pht $0.037$ \pht &\pht $2.9\times 10^{-4}$ \pht & \pht $7.9\times 10^{-3}$ \pht & \pht $9.0\times 10^{-4}$  \pht & \pht $0.069$ \pht  & \pht $1.5\times 10^{-3}$\pht & \pht 0 \pht & \pht  $1.6\times 10^{-6}$ \pht & $0$\pht \\
\hline
$h_3\simeq A$\pht &\pht $0.73$ \pht& \pht $0.076$ \pht&\pht $0.089$ \pht & \pht $2.6\times 10^{-4}$ \pht &\pht $4.3\times 10^{-3}$ \pht & \pht $6.1\times 10^{-4}$ \pht  &\pht  $0.061$ \pht &\pht  $0.037$ \pht  & \pht 0 \pht  &\pht $0$ \pht &\pht  $0$ \pht \\
\end{tblr}
\end{center}
\caption{\justifying \small Branching ratios of the neutral scalars $h_2$ and $h_3$ (with $m_{h_3}=95$~GeV), for the choice of parameters given by the benchmark point exhibited in Table~\ref{tab::95} with $\Im\overline{Z}_7= 0.4$ and $\Re a_U=-0.01$.  The branching ratios for the $Z h_i$ mode correspond to $h_2\to Zh_3$ and $h_3\to Zh_2$, respectively.  Branching ratios less than $10^{-7}$ are indicated by a zero entry above. }
\label{tab::BR95}
\end{table*}

Concerning the 95\,GeV diphoton excess\footnote{The ditau excess of CMS~\cite{CMS:2022goy} at around 100\,GeV is not seen by ATLAS~\cite{ATLAS:2022yrq} and also not confirmed by the $b$-associated channel of CMS. Therefore, we will disregard the ditau channel here.} the combination of ATLAS~\cite{ATLAS:2024bjr} and CMS~\cite{CMS:2024yhz} data prefers a signal strength for the 95\,GeV scalar $S$ of~\cite{Biekotter:2023oen}
\begin{equation} \label{eq::95yy}
\mu_{\gamma\gamma,95}^{\text{LHC}}=\frac{\sigma^{\text{NP}}(pp\to S_{95} \to \gamma\gamma)}{\sigma^{\text{SM}}(pp \to h_{95} \to \gamma\gamma)}=0.24^{+0.09}_{-0.08}\,,
\end{equation}
where $h_{95}$ represents a (hypothetical) SM-like Higgs boson with a mass of 95\,GeV, used to illustrate the (size) of the excess and NP stands for new physics. It has been shown that only small regions in parameter space of the 2HDM with a $\mathbb{Z}_2$ symmetry\footnote{The generic 2HDM with a sizable Yukawa coupling of the top quark to $\mathcal{H}_2$ can explain the 95\,GeV diphoton excess~\cite{Crivellin:2017upt}.} can explain the 95\,GeV excess if this scalar is CP-even~\cite{Haisch:2017gql,Azevedo:2023zkg,Benbrik:2022azi,Benbrik:2024ptw}, whereas a CP-odd solution is even more difficult. The LEP collider found a $2.3\sigma$ excess for a scalar $H$ in $e^+e^-\to Z^*\to Z H$~\cite{LEPWorkingGroupforHiggsbosonsearches:2003ing}, which is most pronounced in $H\to b\bar b$, resulting~\cite{Cao:2016uwt} in~\!\footnote{We rounded the numbers for the LEP signal strength to one significant digit and adjusted the error to recover the $2.3\sigma$ excess reported by LEP.}
\begin{equation}
\mu_{b\bar{b}}^{\text{LEP}}\!=\!\frac{\sigma^{\text{NP}}(e^+e^-\!\!\to\! Z S_{98})}{\sigma^{\text{SM}}(e^+e^-\!\!\to\! Z h_{98})}\frac{\text{BR}(S_{98}\to b\bar{b})}{\text{BR}(h_{98}\to b\bar{b})}\approx0.12\pm 0.06\,,
\end{equation}
where the labels are as in~Eq.(\ref{eq::95yy}). 

\begin{figure}
    \centering
    \includegraphics[width=1\linewidth]{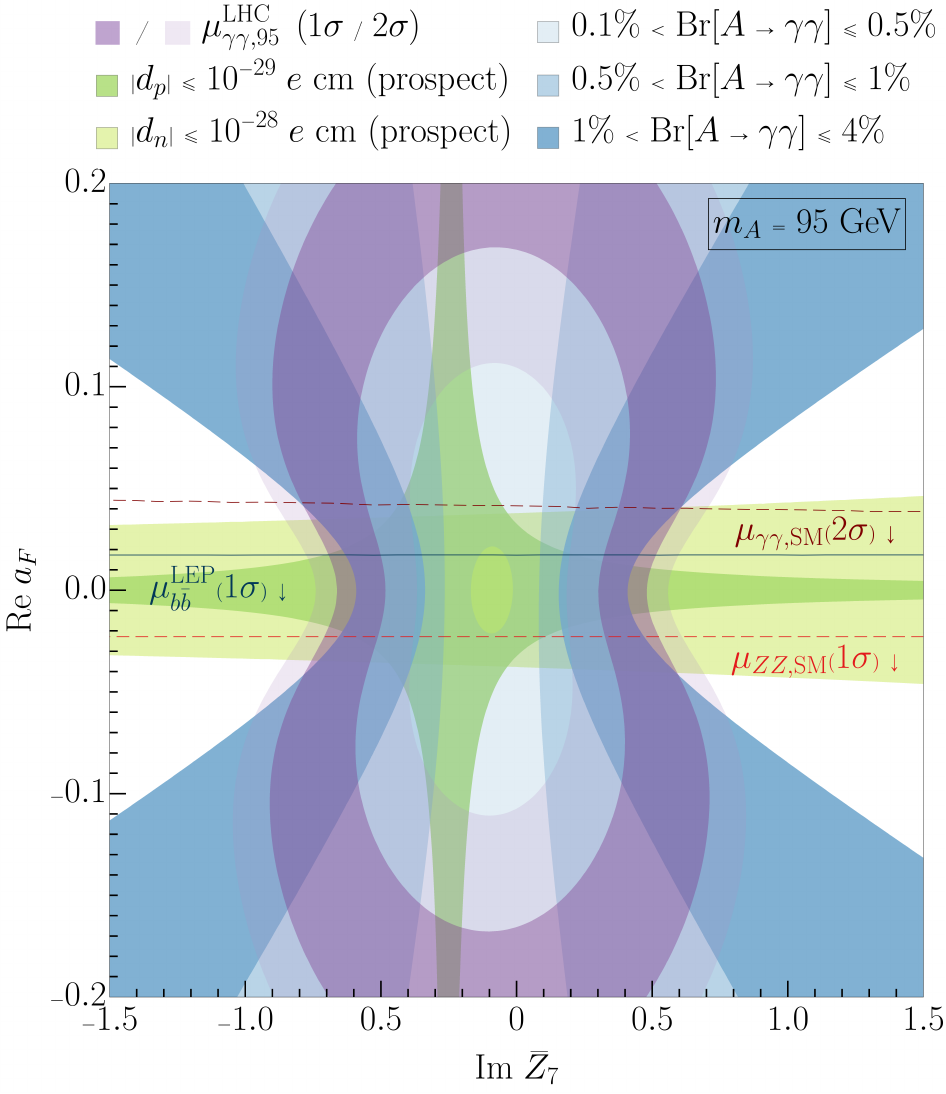}
    \caption{\justifying \small Preferred regions in the $\Im\overline{Z}_7$\,--\,$\Re a_U$ plane from the diphoton excesses at 95\,GeV (blue) and the estimated sensitivity of future neutron and proton EDM measurements (green). The regions above the dashed lines are excluded by the SM Higgs signal strength in $h_1\to\gamma\gamma/ZZ$, while the one below the solid line is preferred by the LEP (98\,GeV) signal strength. The benchmark point exhibited in Table~\ref{tab::95} fixes the other model parameters.}
    \label{fig::95}
\end{figure}

Here we want to consider the option that the diphoton excess is due to the (mostly) CP-odd scalar $h_3\simeq A$. While the LEP signal cannot be explained at the same time by $h_3$, the excess is in fact most pronounced at $\approx$~98\,GeV, indicating that it could be due to another state, which we shall identify as the mostly CP-even scalar $h_2\simeq H$. Note that for the sizable mixing angles preferred by LEP ($\theta_{12}\approx 0.3$), $h_2$ decays dominantly to $b\bar b$ with a small branching ratio to photons such that LHC bounds are not relevant, as indicated in Table~\ref{tab::BR95}.
We have two main production mechanisms for $h_2$ and $h_3$: Drell-Yan (DY) ($pp\to Z^*\to h_2 h_3$, $pp\to W^*\to h_{2,3}H^\pm$) and gluon fusion.\footnote{The gluon-fusion cross section is obtained from rescaling the SM~\cite{LHCHiggsCrossSectionWorkingGroup:2016ypw}, including a factor of $\approx$1.5 due to the axial coupling~\cite{Dawson:1998py,Spira:2016ztx}. For the Drell-Yan production cross section, we used {\tt MadGraph5aMC@NLO}~\cite{Alwall:2011uj,Alwall:2014hca}, including the next-to-next-to leading log (NNLL) and NLO QCD correction factor of $\approx$~1.15 of Refs.~\cite{Ruiz:2015zca,AH:2023hft}.} In the limit of small mixing angles we have
\begin{align}
    \nonumber &\sigma_{\text{\tiny GF}}(pp\to h_3) \approx \frac{1.5}{1+s_{12}^2} \sigma_{\text{\tiny GF}}(pp\to h_{95}) \approx |a_U|^2 100\,\text{pb}  \\
    \nonumber &\sigma_{\text{\tiny DY}}(pp\to h_3 H^\pm) \approx \sigma_{\text{\tiny DY}}(pp\to h_2 H^\pm) \approx 0.31\,\text{pb} \\[.2cm]
    &\sigma_{\text{\tiny DY}}(pp\to h_3 h_2) \approx 0.28\,\text{pb}
\end{align}
where we have taken 130\,GeV for the charged Higgs mass in light of the ATLAS excess in $t\to (H^\pm\to cb) b$~\cite{ATLAS:2023bzb}. For simplicity, let us consider the dependence on $a_U$ and $\Im\zvii$ while setting $a_D=a_E=0$, which strongly suppresses an effect in the very constraining electron EDM. 
We fix the other relevant model parameters as indicated in Table~\ref{tab::95}.   The branching ratios of the $h_2$ and $h_3$ decay modes corresponding to this benchmark point are exhibited in Table~\ref{tab::BR95}.
\captionsetup[table]{
  position=above, aboveskip=4pt,
  belowskip=-4pt
}
\begin{table}[t!]
\begin{center}
\SetTblrInner{colsep=4pt}
\begin{tblr}{c|c|c|c|c|c|c|c|c}
$m_{h_1}$	&	$m_{h_2}$	&	$m_{h_3}$	&	$m_{H^\pm}$		&	$\theta_{12}$	&	$\theta_{13}$	&	$Z_2$	&	$Z_3$	&	$\Re\zvii$ \\
\hline
$125$	&	$98$	&	$95$	&	$130$	&	$0.25$	&	$0.01$	&	$0.2$		&	$-0.2$	&	$0.1$
\end{tblr}
\end{center}
\caption{\justifying \small Benchmark point used for the interpretation of the $\gamma\gamma$ excesses at 95\,GeV and the $b\bar b$ excess at 98\,GeV. We fixed $a_D=a_E=0$ and Arg~$a_U=-0.03$, and masses are given in units of GeV.}
\label{tab::95}
\end{table}
In Fig.~\ref{fig::95}, we show the
preferred regions in the $\Im\overline{Z}_7$\,--\,$\Re a_U$ plane.\footnote{Note that charged Higgs boson searches are not constraining for our setup~\cite{ATLAS:2024oqu} and we checked with \texttt{HiggsTools}~\cite{Bahl:2022igd} that also no other search channels implemented there are violated for this benchmark point. Additionally, for the parameter space explaining the excesses without violating other bounds, we checked for consistency with vacuum stability and perturbative unitarity.}  Note the complementary between the proton and the neutron EDM.
In particular, future EDM measurements can cover most of the parameter space in which the 95\,GeV diphoton excess is explained. 

\subsection{152\,GeV excess}
\captionsetup[table]{
  position=above, aboveskip=4pt,
  belowskip=-4pt
}
\begin{table*}
\begin{center}
\SetTblrInner{colsep=.5pt}
\begin{tblr}{c|c|c|c|c|c|c|c|c|c|c}
	&	$b\bar b$		& 	$\tau^+\tau^-$	&	$c\bar c$	&	$W^+ W^-$	&	$ZZ$	 	&	$gg$		&	$\gamma\gamma$	&	$Zh_1$	&	$Zh_i$	&\pht 	$W^\pm H^\mp$\\
\hline
$h_2 \simeq H$ \pht & \pht $2.7\times 10^{-4}$ \pht & \pht $3.3\times 10^{-5}$ \pht &\pht  $1.4\times 10^{-5}$ \pht & \pht $0.021$ \pht & \pht $7.2\times 10^{-3}$ \pht & \pht $1.5\times 10^{-4}$ \pht & \pht $8.5\times 10^{-7}$ \pht & \pht $4.7\times 10^{-7}$ \pht & \pht $0.045$ \pht & \pht $0.93$ \\
\hline
$h_3 \simeq A $ \pht  & \pht $0.026$ \pht & \pht $2.7\times 10^{-3}$ \pht & \pht $1.6\times 10^{-3}$ \pht &\pht $1.3\times 10^{-3}$ \pht & $\pht 1.4\times 10^{-4}$ \pht & \pht $1.3\times 10^{-4}$ \pht & \pht $0.012$ \pht & \pht $1.0\times 10^{-4}$ \pht & \pht $0$ \pht & \pht $0.95$ \\
\end{tblr}
\end{center}
\caption{\justifying \small Branching ratios of the neutral scalars $h_2$ and $h_3$ (with $m_{h_3}=152$~GeV), for the choice of parameters given by the benchmark point exhibited in Table~\ref{tab::152} with $\Im\overline{Z}_7= 0.8$ and $\Re a_F=0.01$. The branching ratios for the $Z h_i$ mode correspond to $h_2\to Zh_3$ and $h_3\to Zh_2$, respectively. Branching ratios less than $10^{-7}$ are indicated by a zero entry above.}
\label{tab::BR152}
\end{table*}

\begin{figure}[h!]
    \includegraphics[width=1\linewidth]{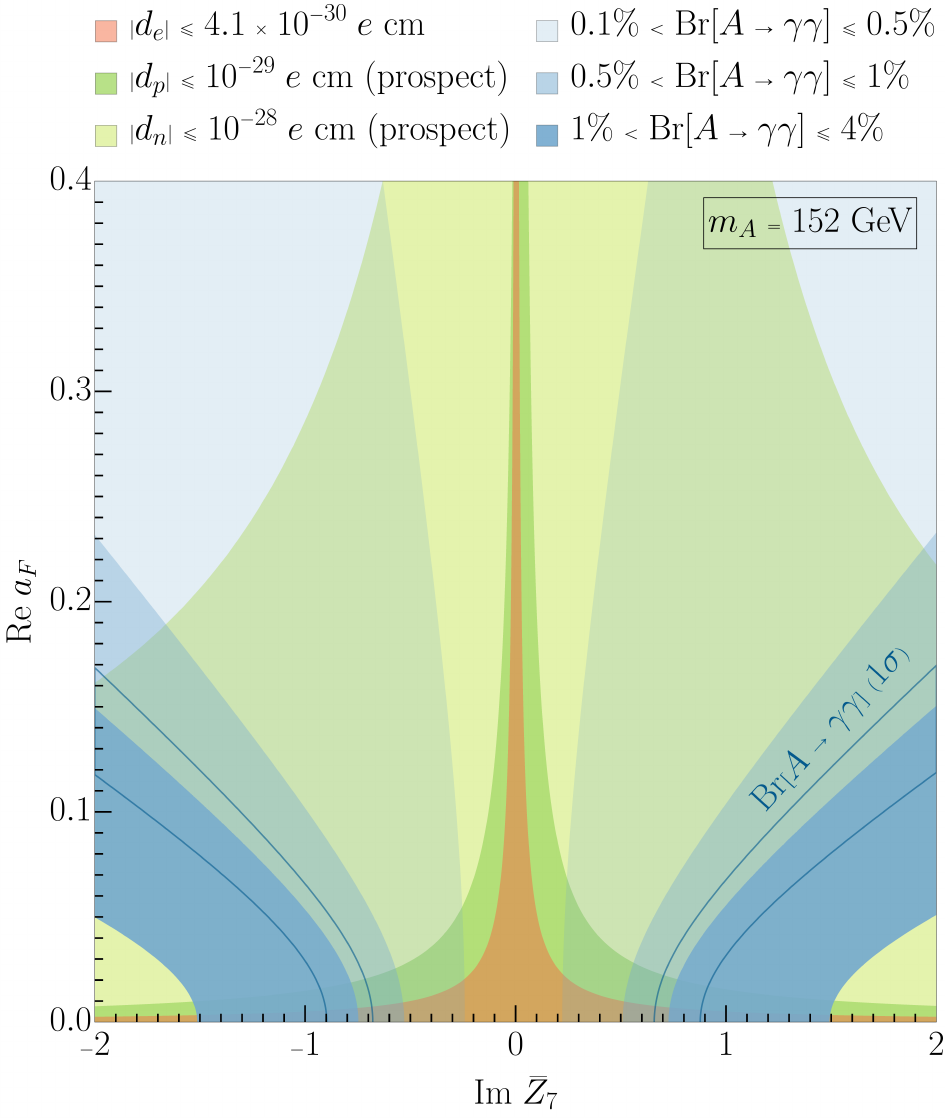}
    \caption{\justifying \small BR$(h_3 \to \gamma \gamma)$ in the $\Im\overline{Z}_7$\,--\,$\Re a_F$ plane as well as the allowed region from the electron EDM (orange) and the regions where future neutron and proton EDM measurements will be sensitive (green). The preferred $1\sigma$ region from the $\gamma\gamma+X$ excesses at 152\,GeV is indicated by the band with dark blue solid lines. Constraints from $h_1\to\gamma\gamma/ZZ$ signal strength are satisfied within the whole depicted area at the $1\sigma$ and thus not displayed. The benchmark point exhibited in Table~\ref{tab::152} fixes the other model parameters.}
    \label{fig::152}
\end{figure}

Here, the relevant production process is Drell-Yan as we are considering the associated production of the 152\,GeV boson, i.e.~$\gamma\gamma+X$ which is significantly more sensitive to physics beyond the SM than the inclusive measurements. From the analysis of Ref.~\cite{Banik:2024ftv},\footnote{Although Ref.~\cite{Banik:2024ftv} considered the case in which $h_2\simeq H$ is the 152\,GeV candidate, since only branching ratios of $H^\pm$ matter and the branching ratio of the neutral scalar to photons is the fit parameter, the results apply to our case.}  \pagebreak
we see that the preference for a nonzero diphoton branching ratio with a best-fit value of $\approx 1.3\%$ is greater than $4\sigma$. 
Here, we consider for simplicity the case in which the relative coupling strength to all fermions is the same, $a_U=a_D=a_E=a_F$. 
The branching ratios of the neutral scalars $h_2$ and $h_3$ (for $m_{h_3}=152$~GeV) are given in Table~\ref{tab::BR152} for the benchmark point specified in
Table~\ref{tab::152}.\footnote{In contrast to the benchmark point of Section~\ref{ninefive} where $a_D=0$, here we have assumed that both $a_U$ and $a_D$ are nonzero.   We have checked that values of the charged Higgs mass as low as $m_{H^\pm}=130$~GeV are not excluded by the observed rate for $b\to s\gamma$, where the dominant new physics contribution (via a charged Higgs loop) is proportional to $\Re(a_D^* a_U)$.  Although such low charged Higgs masses are untenable in the Type-II 2HDM~\cite{Misiak:2020vlo}, the corresponding constraints in the flavor-aligned 2HDM are considerably weaker in a large region of its parameter space~\cite{Connell:2023jqq,Coutinho:2024zyp}.
In particular, with the benchmark parameters given in Table~\ref{tab::152}, the entire plane shown in Fig.~\ref{fig::152} is allowed.}
\captionsetup[table]{
  position=above, aboveskip=4pt,
  belowskip=-4pt
}
\begin{table}[t!]
\begin{center}
\SetTblrInner{colsep=4pt}
\begin{tblr}{c|c|c|c|c|c|c|c|c}
$m_{h_1}$	&	$m_{h_2}$	&	$m_{h_3}$	&	$m_{H^\pm}$		&	$\theta_{12}$	&	$\theta_{13}$	&	$Z_2$	&	$Z_3$	&	$\Re\zvii$ \\
\hline
$125$	&	$200$	&	$152$	&	$130$	&	$0.01$	&	$0.001$	&	$0.2$	&	$-0.2$	&	$0.1$
\end{tblr}
\end{center}
\caption{\justifying \small Benchmark point used for the interpretation of the $\gamma\gamma+X$ excesses at 152\,GeV. We fixed $a_U=a_D=a_E=a_F$ with Arg~$a_F=-0.01$, and masses are given in units of GeV.}
\label{tab::152}
\end{table}
The comparison of the preferred diphoton branching ratios as well as the bounds from EDMs are shown\footnote{We checked with \texttt{HiggsTools}~\cite{Bahl:2022igd} that the parameter space addressing the observed excesses is consistent with other direct searches, SM Higgs signal strength and electroweak precision data and satisfies vacuum stability and perturbative unitarity.} in Fig.~\ref{fig::152}. One can see that the electron EDM enforces the product of $\Im\zvii\times |a_F|$ to be small, such that the 152\,GeV excess can only be explained for small values of the couplings $|a_F|$, corresponding to the region where DY is the main production mechanism. If $a_E=0$, then the electron EDM constraint would be avoided, but the scenario could still be tested by future neutron and proton EDM measurements. 
In light of the large branching ratio for $A\to W^\pm H^\mp$ (with, say, the $W$ boson off-shell) as shown in Table~\ref{tab::BR152}, searches for the charge Higgs boson in future LHC runs will provide an important test of this scenario.

\section{Conclusions}
\label{conclude}

In this paper, we proposed that a large branching ratio to photons of the (mostly) CP-odd scalar $h_3\simeq A$ in the flavor-aligned 2HDM can be achieved if the Yukawa flavor-alignment parameters are small and the parameter $ \zvii$ has a sizable imaginary part. This then acts as a source of CP violation, giving rise to nonvanishing EDMs of fundamental fermions. We considered two benchmark points motivated by the diphoton excesses at 95\,GeV and 152\,GeV, explored the correlations with the electron EDM, and noted that these regions of parameter space can be tested by future neutron and proton EDM experiments. Moreover, we found that if $h_3\simeq A$ accounts for the 95\,GeV $\gamma\gamma$ excess, then $h_2\simeq H$ could explain the LEP excess in Higgs-strahlung that is more pronounced at 98\,GeV than at 95\,GeV.

\acknowledgements

A.C.~thanks Martin Hoferichter for useful discussion on the neutron and proton EDMs. G.C.~thanks Wolfgang Altmannshofer for his clarification on the code used in the EDMs calculations. 
A.C.~acknowledges support from a professorship grant from the Swiss National Science Foundation (No.\ PP00P21\_76884).  H.E.H.~is supported in part by the U.S.~Department of Energy Grant No.~\uppercase{DE-SC}0010107. 

\appendix

\section{YUKAWA SECTOR OF THE 2HDM} \label{app::Yukawa}
In the Higgs basis, the 2HDM Yukawa coupling Lagrangian
is given by \eq{yukawas}.  First, we focus on the terms of the Yukawa Lagrangian
involving the neutral scalar fields.  It then follows that
 \beqa 
 -\mathscr{L}_{Y}&=&(\boldsymbol{\wh{\kappa}_U})_{mn}
\mathcal{H}_1^{0\dagger}\, \whbar{u}_{mL} \wh{u}_{nR}
+(\boldsymbol{\wh{\rho}_U})_{mn}
\mathcal{H}_2^{0\dagger}\, \whbar{u}_{mL} \wh{u}_{nR}
\nonumber \\
&&
+ (\boldsymbol{\wh{\kappa}_D^{\dagger}})_{mn}
\mathcal{H}_1^0\,\whbar{d}_{mL} \wh{d}_{nR}
+ (\boldsymbol{\wh{\rho}_D^{\dagger}})_{mn}
\mathcal{H}_2^0\,\whbar{d}_{mL} \wh{d}_{nR}
\nonumber \\[4pt]
&& 
+(\boldsymbol{\wh{\kappa}_D^{\dagger}})_{mn}\mathcal{H}_1^0\,\whbar{e}_{mL} \wh{e}_{nR}
+ (\boldsymbol{\wh{\rho}_E^{\dagger}})_{mn}
\mathcal{H}_2^0\,\whbar{e}_{mL} \wh{e}_{nR} \nonumber \\[4pt]
&& +{\rm H.c.}
\eeqa
where $f_{R,L}\equiv\half(1\pm\gamma\lsub{5})f$, with $f=u,d,\nu,e$ and
there is an implicit sum over the repeated fermion generation indices $m$, $n\in\{1,2,3\}$. The hatted fields correspond to interaction eigenstates.  
Setting 
$\mathcal{H}_1^0=\mathcal{H}_1^{0\,\dagger}=v/\sqrt{2}$ yields the fermion mass matrices
\beqa
(\boldsymbol{\widehat{M}_U})_{mn} &=& \frac{v}{\sqrt{2}} (\boldsymbol{\widehat{\kappa}_U})_{mn}\,,\\
(\boldsymbol{\widehat{M}_{D,E}})_{mn} &=& \frac{v}{\sqrt{2}} (\boldsymbol{\widehat{\kappa}_{D,E}^{\dagger}})_{mn}\,,
\eeqa
and the neutrino mass matrix $\boldsymbol{\widehat{M}_N}=0$. The singular value decompositions of $\boldsymbol{\widehat{M}}_U$ and $\boldsymbol{\widehat{M}}_D$ yield
\beq \label{svd1}
L_u^{\dagger} \,\boldsymbol{\widehat{M}_U}\, R_u\equiv \boldsymbol{{M}_U}\,,\qquad
L_d^{\dagger} \,\boldsymbol{\widehat{M}_D} \,R_d\equiv \boldsymbol{{M}_D}
\eeq
where $\boldsymbol{{M}_U}$ and $\boldsymbol{{M}_D}$ are diagonal up- and down-type quark mass matrices with real and nonnegative diagonal elements,
and the unitary matrices $L_f$ and $R_f$ ($f=u,d$) relate hatted interaction-eigenstate fermion fields with unhatted mass-eigenstate fields, 
\beq
\widehat f_{mL} = (L_f)_{mn} f_{nL}\,,\qquad\!\! \widehat f_{mR}=(R_f)_{mn} f_{nR}\,.
\eeq
The Cabibbo-Kobayashi-Maskawa (CKM) matrix is denoted by
$\boldsymbol{K}\equiv L_u^\dagger L_d$.  

Likewise, the singular value decomposition of $\boldsymbol{\widehat{M}_E}$ yields 
\beq \label{svd2}
L_e^{\dagger} \,\boldsymbol{\widehat{M}_E} \,R_e\equiv \boldsymbol{{M}_E}\,,
\eeq
where $\boldsymbol{{M}_E}$ is the diagonal charged lepton mass matrix with real and nonnegative diagonal elements, and
the mass-eigenstate lepton fields are given (for $f=\nu$, $e$) by
\beq
\widehat f_{mL} = (L_e)_{mn} f_{nL}\,,\qquad\!\! \widehat f_{mR}=(R_e)_{mn} f_{nR}\,.
\eeq

The physical (basis-invariant) $\rho$-type Yukawa couplings are complex \mbox{$3\times 3$} matrices, 
\beqa
\boldsymbol{\rho_{U}}&\equiv & L_u^{\dagger} \,\boldsymbol{\wh{\rho}_{U}}R_u\,,\qquad \quad  
\boldsymbol{\rho_D^{\dagger}}\equiv  L_d^{\dagger} \,\boldsymbol{\wh{\rho}_{D}^\dagger}R_d\,,\nn \\
&& \qquad  \,\,\, \boldsymbol{\rho_E^{\dagger}}\equiv L_e^{\dagger}\, \boldsymbol{\wh{\rho}_E^{\dagger}} R_e\,, \label{physicalrho}
\eeqa
that generically yield off-diagonal neutral Higgs--fermion interactions.
The corresponding neutral Higgs--fermion interactions involving mass-eigenstate 
scalar and fermion fields can now be obtained.  If we additionally include the charged Higgs--fermion interactions starting with \eq{yukawas} and replace the interaction-eigenstate fields with mass-eigenstate fields, we end up with:
\begin{widetext}
\beqa
 -\mathscr{L}_Y &&\,=\,\overline{U} \biggl\{\frac{\boldsymbol{M_U}}{v}q_{k1}+\frac{1}{\sqrt{2}}
\Bigl[q^*_{k2}\,\boldsymbol{\rho_U} \mathcal{P}_R+
q_{k2}\,\boldsymbol{\rho_U^{\dagger}} \mathcal{P}_L\Bigr]\biggr\}U h_k 
+ \overline{D} \biggl\{\frac{\boldsymbol{M_D}}{v}q_{k1}+\frac{1}{\sqrt{2}}\Bigl[q_{k2}\,\boldsymbol{\rho_D^{\dagger}} \mathcal{P}_R+q^*_{k2}\,\boldsymbol{\rho_D} \mathcal{P}_L\Bigr]\biggr\}Dh_k \nonumber \\[5pt]
&& \!\!\! \!\!\!+\, \overline{E} \biggl\{\frac{\boldsymbol{M_E}}{v}q_{k1}+\frac{1}{\sqrt{2}}\Bigl[q_{k2}\,\boldsymbol{\rho_E^{\dagger}} \mathcal{P}_R+q^*_{k2}\,\boldsymbol{\rho_E} \mathcal{P}_L\Bigr]\biggr\}Eh_k +\biggl\{\overline{U}\Bigl[\boldsymbol{K\boldsymbol{\rho_D^{\dagger}}}\mathcal{P}_R
-\boldsymbol{\rho_U^{\dagger} K}\mathcal{P}_L\Bigr] D H^+ +\overline{N}\boldsymbol{\rho_E^{\dagger}}\mathcal{P}_R EH^++ {\rm H.c.}\biggr\}\,, \nn \\
\phantom{line} \label{YUK2}
\eeqa
\end{widetext}
with an implicit sum over the index $k=1,2,3$,
where $\mathcal{P}_{R,L}\equiv \half(1\pm\gamma\lsub{5})$
and the mass-eigenstate fields of the down-type quarks, up-type quarks, charged leptons, and neutrinos are denoted by
$D=(d,s,b)^{\T}$, $U\equiv (u,c,t)^{\T}$, $E=(e,\mu,\tau)^{\T}$, and $N=(\nu_e,\nu_\mu,\nu_\tau)^{\T}$, respectively.

As previously noted, the matrices $\boldsymbol{\rho_F}$ are in general complex and nondiagonal, which can generate dangerously large tree-level flavor-changing neutral currents (FCNCs) mediated by neutral scalars. In this work, we have employed the flavor-aligned 2HDM~\cite{Pich:2009sp,Pich:2010ic,Eberhardt:2020dat,Choi:2020uum,Lee:2021oaj,Serodio:2011hg,Cree:2011uy,Connell:2023jqq} in which the $\boldsymbol{\rho_F}$ are proportional to the corresponding diagonal fermion mass matrices $\boldsymbol{M_F}$ without imposing any symmetry (such as the $\mathbb{Z}_2$ symmetry used in constructing the Type I, II, X, or Y 
2HDM~\cite{Hall:1981bc,Barger:1989fj,Aoki:2009ha}, which naturally yields flavor-diagonal neutral scalar couplings). 
In particular, we define the \textit{flavor-alignment parameters} $a_F$ via $\boldsymbol{\wh{\rho}_F}= a_F\boldsymbol{\wh{\kappa}_F}$ for $F=U,D,E$, where the (potentially) complex numbers $a_F$ are invariant under a scalar field basis transformation.
In light of \eqss{svd1}{svd2}{physicalrho}, it follows that
\beq \label{aligned}
\boldsymbol{\rho_F}=\frac{\sqrt{2}}{v} a_F \boldsymbol{M_F}\,,\qquad\quad \text{for $F=U,D,E$}.
\eeq
\Eqs{YUK2}{aligned} then yield the Higgs--fermion Yukawa couplings of the flavor-aligned 2HDM, 
\begin{widetext}
\beqa
 -\mathscr{L}_Y &=& \frac{1}{v}\,\overline U \boldsymbol{M_U}\sum_{k=1}^3 \bigl(q_{k1}
+ q_{k2}^* a_U \mathcal{P}_R+q_{k2} a_U^*\mathcal{P}_L\bigr)Uh_k  
 +\frac{1}{v}\sum_{F=D,E} \overline F  \boldsymbol{M_F}\sum_{k=1}^3 \bigl(q_{k1}
+ q_{k2} a_F^* \mathcal{P}_R+q^*_{k2}a_{F}\mathcal{P}_L\bigr) Fh_k \nn \\[5pt]
 && +\frac{\sqrt{2}}{v}\biggl\{\overline U\bigl[a_D^*\boldsymbol{KM_D}\mathcal{P}_R-a_U^*\boldsymbol{M_U K}\mathcal{P}_L\bigr] DH^+ 
 +a_E^*\overline N\boldsymbol{M_E} \mathcal{P}_R EH^+
+{\rm H.c.}\biggr\}. \label{YUK4}
\eeqa
\end{widetext}

\section{LOOP FUNCTIONS}
\label{app::LoopA}
We list the loop functions used in the computations of the partial decay widths of the scalars to two photons~\cite{Gunion:1989we}:
\begin{align}
    A_W(\tau) &= 2 + 3\tau + 3\tau (2-\tau)g(\tau) \\[.2cm]
    A_{H^\pm}(\tau) &= \tau [1-\tau g(\tau)] \\[.2cm]
    A_f^{0}(\tau) &= -2 \tau [1+(1-\tau)g(\tau)] \\[.2cm]
    A_f^{5}(\tau) &=  -2\tau g(\tau)
\end{align}
where the function $g(\tau)$ is given by
\begin{equation}
    g(\tau) = \begin{cases} 
     \phantom{-\frac{1}{4} }
    \left[\sin^{-1}(\sqrt{1/\tau})\right]^2\,, & \text{for $\tau \ge 1$}, 
    \\[10pt]   
    -\frac{1}{4} \left[\log{\displaystyle\left(\frac{1+\sqrt{1-\tau}}{1-\sqrt{1-\tau}}\right)} - i\pi\right]^2\,, & \text{for $\tau < 1$}. 
\end{cases} 
\end{equation}
\phantom{xxxx} \\ \phantom{xxxx}  \\ \phantom{xxxx}

\bibliographystyle{utphys}
\bibliography{apssamp}
\end{document}